\documentclass{emulateapj}

\usepackage{graphicx}

\shorttitle{Spiderweb Galaxy}
\shortauthors{Miley et al.}

\begin{document}

\title{The Spiderweb galaxy: A forming massive cluster galaxy at $z\sim2$}

\author{George K. Miley\altaffilmark{1}, Roderik A. Overzier\altaffilmark{1}, Andrew W. Zirm\altaffilmark{2}, Holland C. Ford\altaffilmark{2}, Jaron Kurk\altaffilmark{3}, Laura Pentericci\altaffilmark{4}, John P. Blakeslee\altaffilmark{5}, Marijn Franx\altaffilmark{1}, Garth D. Illingworth\altaffilmark{6}, Marc Postman\altaffilmark{7}, Piero Rosati\altaffilmark{8}, Huub J.A. R{\"o}ttgering\altaffilmark{1}, Bram P. Venemans\altaffilmark{9} and Eveline Helder\altaffilmark{1}}

\altaffiltext{1}{Leiden Observatory, University of Leiden, P.O. Box 9513, Leiden, 2300 RA, Netherlands}

\altaffiltext{2}{Department of Physics and Astronomy, Johns Hopkins University, Baltimore, MD 21218}
\altaffiltext{3}{Max Planck Institut f\"ur Astronomie, Heidelberg, Germany}
\altaffiltext{4}{INAF-Osservatorio di Roma, Sede di Monteporzio Catone, Via di Frascati, 33, Rome I-00040, Italy}
\altaffiltext{5}{Department of Physics and Astronomy, Washington State University, Pullman, WA 99164-2814USA}
\altaffiltext{6}{Lick Observatory, University of California, Santa Cruz, CA 95064}
\altaffiltext{7}{Space Telescope Science Institute, 3700 San Martin Drive, Baltimore, MD 21218}
\altaffiltext{8}{European Southern Observatory, Karl-Schwarzschild-Strasse 2, D-85748, Garching, Germany}
\altaffiltext{9}{Institute for Astronomy, Madingley Road, Cambridge CB3 0HA, UK}
\email{miley@strw.leidenuniv.nl}

\begin{abstract}

We present a deep image of the radio 
galaxy MRC 1138--262 taken with the {\it Hubble Space Telescope (HST)} at a redshift of $z = 2.2$. The galaxy is known to have properties of a cD galaxy progenitor 
and be surrounded by a 3 Mpc-sized structure, identified with a protocluster. The morphology shown on 
the new deep {\it HST} ACS image is reminiscent of a spider's web. More than 10 
individual clumpy features are observed, apparently star-forming satellite galaxies in the process of merging with the 
progenitor of a dominant cluster galaxy 11 Gyr ago. There is an extended emission component, implying 
that star formation was occurring over a 50$\times$40 kpc region at a 
rate of more than 100 $M_{\sun}$ yr$^{-1}$. A striking feature of the newly named 
``Spiderweb galaxy'' is the presence of several faint linear galaxies 
within the merging structure. The dense environments and fast galaxy motions 
at the centres of protoclusters may stimulate the formation of these 
structures, which dominate the faint resolved galaxy populations in the Hubble Ultra Deep Field. 
The new image provides a unique testbed for simulations of forming dominant cluster galaxies.
\end{abstract}

\keywords{galaxies: clusters: general -- galaxies: cD -- galaxies: high redshift -- galaxies: active}

\section{Introduction}

Distant powerful radio galaxies are important laboratories for studying the 
formation and evolution of massive galaxies, because they are among the most 
luminous and largest known galaxies in the early Universe 
and likely progenitors of dominant cluster galaxies 
\citep[e.g.][]{mil2000}. 

They are generally embedded in giant (cD-sized) ionized gas halos \citep[e.g.][]{vano1997}
surrounded by galaxy overdensities, whose structures have sizes of a few Mpc \citep{pen2000,ven2002,ven2005}. 
The radio galaxy hosts have clumpy optical morphologies \citep{pen1998,pen1999}, spectra indicative of 
extreme star formation \citep[e.g.][]{dey1997}, and large stellar masses \citep{vil2006}.
Because the radio lifetimes (few times 10$^{7}$ yr) are much smaller than cosmological 
timescales, the statistics are consistent with every dominant cluster galaxy 
having gone through a luminous radio phase during its evolution \citep{ven2002}. Hence, 
distant radio galaxies may be typical progenitors of galaxies that dominate the cores of local clusters.

The radio source MRC 1138--262, identified with a galaxy at $z=2.168$, is one 
of the most intensively studied distant radio galaxies \citep{pen1997,pen1998}. 
Several of its properties are those expected of 
the progenitor of a dominant cluster galaxy. 
The $K$-band luminosity corresponds to a stellar mass of $\sim$$10^{12}$ $M_{\sun}$ \citep{pen1998}, 
implying that MRC 1138--262 is one of the most massive galaxies 
known at $z>2$. The host galaxy is surrounded by a giant Ly$\alpha$ halo \citep{pen2000,kur2004a} and 
the Faraday rotation of the radio source is among the largest known \citep{car1997},  
indicating that the system is embedded in a dense hot ionized gas with an ordered magnetic field.

The radio galaxy is associated with a 3 Mpc-sized structure of 
galaxies, of estimated mass $\sim $ 2 $\times$$10^{14}$ $M_{\sun}$, the presumed antecedent of a local cluster. 
The presence of this ``protocluster" has 
been deduced using three independent selection techniques. There are overdensities of 
Ly$\alpha$ and H$\alpha$ emission lines objects and galaxies having 4000 \AA\ break 
continuum features at the approximate redshift of the radio galaxy 
\citep{pen2000,kur2004a,kur2004b}.

Previous observations of MRC 1138--262 with the {\it Hubble Space 
Telescope (HST)} indicated that its optical emission is clumpy \citep{pen1998}, 
indicative of a merging structure. Here we present a new 
{\it HST} image of the radio galaxy that reaches 2 
mag fainter than previous images and shows the merging processes in 
unprecedented detail.

Throughout this Letter 
we assume a standard cosmology with $H_0$ = 72 km s$^{-1}$ Mpc$^{-1}$, $\Omega 
_{M}$ = 0.27, and $\Omega _{\Lambda }$ = 0.73, implying that at the 
distance of MRC 1138--262, an angular scale of 1\arcsec\ corresponds to a projected 
linear scale of 8.3 kpc.

\section{Observations}

\begin{figure*}
\begin{center}
\includegraphics[width=\textwidth]{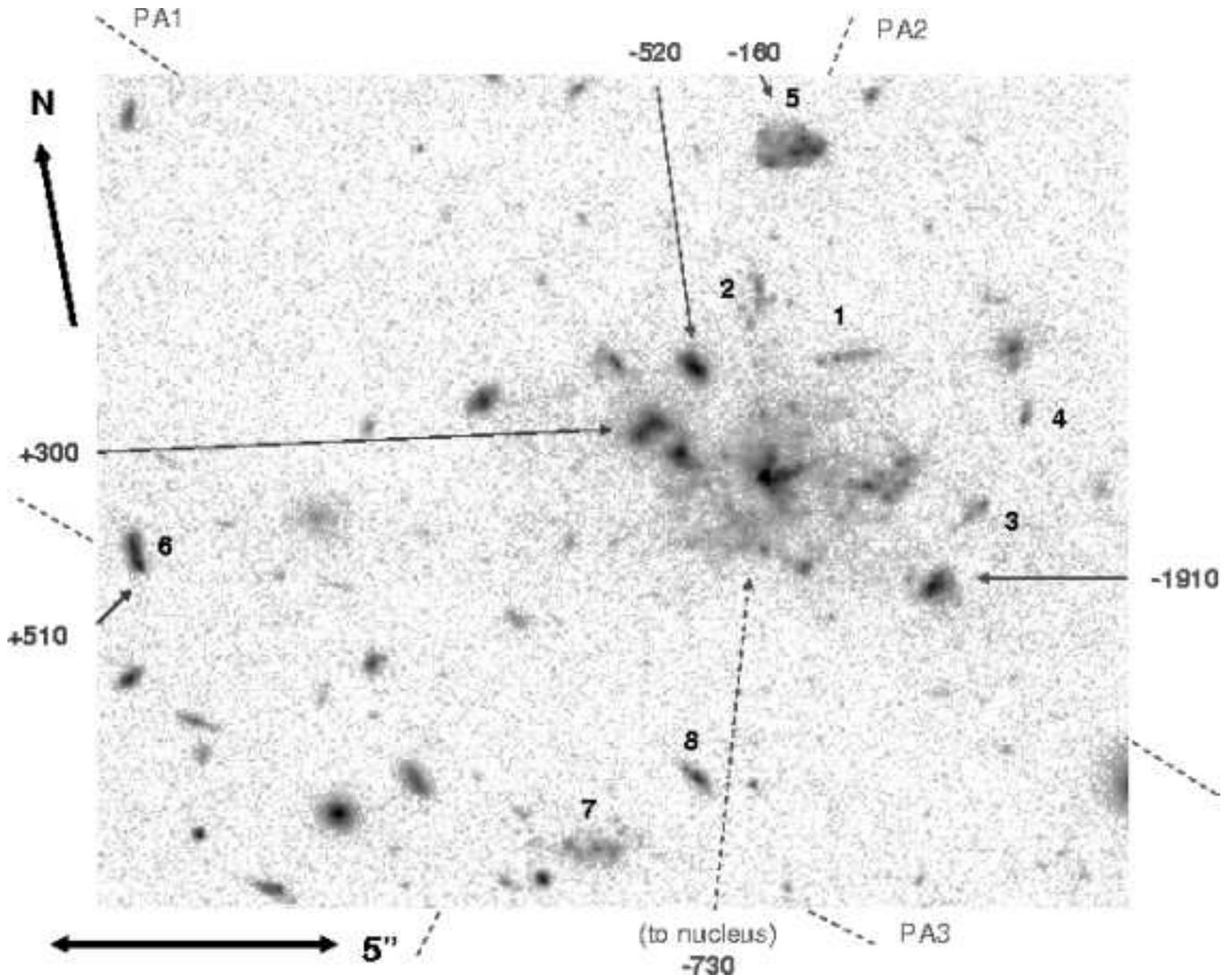}
\end{center}
\caption{
Composite image of a 23\arcsec\ by 18\arcsec\ region at the core of the MRC 1138--262 
protocluster taken with the ACS through the $g_{475}$ + $I_{814}$ filters, using a total 
of 19 orbits. Also shown are hitherto unpublished rest-frame Ly$\alpha$ emission velocities 
in kilometers per second, measured through 1\arcsec\ wide spectrograph slits in three position angles indicated 
by the dashed lines. These were obtained using the FORS spectrograph on the Antu telescope 
of the VLT during 2000 March and April \citep{kur2003}. The velocities were measured at the 
peaks of the Ly$\alpha$ emission profiles and are relative to the median velocity of Ly$\alpha$ 
absorption. Following \citet{kur2003}, the nucleus is taken to be the position of the peak 
H$\alpha$. This coincides with the peak in ACS continuum emission, indicated by the extrapolation 
of the red arrow corresponding to - 730 km s$^{-1}$, the velocity of the nuclear Ly$\alpha$ 
emission. Eight of the satellite galaxies (flies) that have chain, tadpole, or clumpy 
morphologies are indicated by numerals 1--8.}
\end{figure*}

We obtained deep images of MRC 1138--262 with the Advanced Camera for Surveys 
\citep[ACS;][]{for1998} on the {\it HST}. 
The centre of the 1138--262 protocluster was observed with the ACS during  2005 17--22 May in two 3\farcm4$\times$3\farcm4 
ACS fields that overlapped by 1\arcmin\, in a region that includes the radio galaxy. The total exposure time in the overlapping 
region was nine orbits with the F475W ($g_{475}$) filter 
and 10 orbits with the F814W ($I_{814}$) filter.  
The filters were selected to sample the continuum radiation with maximum 
sensitivity and colour discrimination, while minimizing contamination from 
bright emission lines. 
Although the \hbox{C~\footnotesize$\rm IV$} $\lambda$1549 and \hbox{He~\footnotesize$\rm II$} $\lambda$1640 lines fall 
within the $g_{475}$ passband, their measured 
rest-frame equivalent widths are 6 and 10 {\AA}, respectively  \citep{rot1997}, 
implying that their effect on the continuum image is negligible. The 
observations were processed through the ACS GTO pipeline \citep{bla2003}
to produce registered, cosmic-ray rejected images. 
A deep 19 orbit composite image of the continuum emission 
was then produced by adding the $g_{475}$ and $I_{814}$ images. The $2\sigma$ depth in the overlapping region is 29.3 and 29.0 mag in the respective $g_{475}$ and $I_{814}$ images 
measured in a square $0\farcs45$ diameter aperture.

\section{Results}

The {\it HST} image of the host galaxy and its immediate surroundings is 
shown in Figure 1, with rest-frame velocities of the Ly$\alpha$ emission corresponding to several of the ACS continuum clumps.  
Figure 2 illustrates the relation of the continuum 
optical emission to the associated gaseous halo and relativistic plasma. 
The giant 
gaseous halo extends by at least 25\arcsec\ ($\sim$200 kpc) and is one of the 
largest Ly$\alpha$ structures known in the Universe. 

\begin{figure*}
\begin{center}
\includegraphics[width=\textwidth]{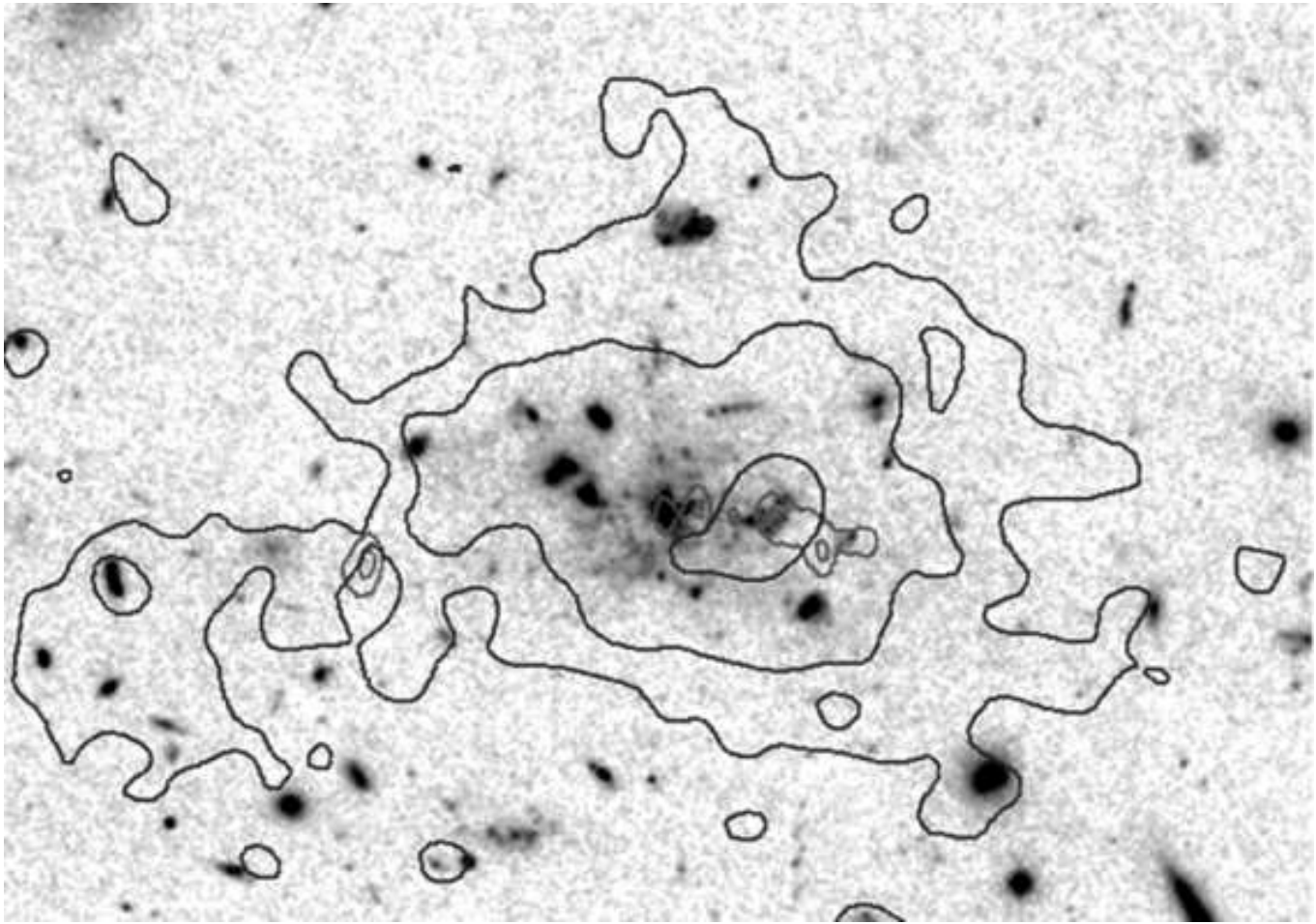}
\end{center}
\caption{
VLT Ly$\alpha $ contours ({\it blue}, resolution $\sim$1\arcsec\ FWHM) delineating 
the gaseous nebula and the VLA 8 GHz contours ({\it red}, resolution 
$\sim0\farcs3$) delineating the nonthermal radio emission are superimposed on the 
composite ($g_{475}$ + $I_{814}$) ACS image. The image shows a 33\arcsec\ by 
23\arcsec\ region rotated 10\degr\ from north. 
The gaseous nebula extends for $>$200 kpc and is comparable in size with the 
envelopes of cD galaxies in the local Universe.}
\end{figure*}

There are several features in the figures that are of interest:

1. The optical continuum emission of the galaxy
consists of at least 10 distinct clumps.
The clumps are presumably satellite 
galaxies that are still merging with MRC 1138--262. They have sizes of 
typically $\sim$0\farcs1--0\farcs5, corresponding to $\sim$1--5 kpc, i.e. 
comparable to the typical sizes of Lyman break galaxies \citep{fer2004,bou2004}.

2. Several of the satellites have elongated structures reminiscent of chain and tadpole 
galaxies recently found to dominate the resolved population of the Hubble Ultra Deep Field 
\citep[HUDF,][]{elm2005,str2006}. Examples are denoted by numbers 1, 3, and 4 in Figure 1. 
Another linear, distorted galaxy is seen 3'' north of the nucleus (2), and several galaxies 
having double (6, 8) or clumpy morphologies embedded in diffuse emission (5, 7) lie at slightly 
larger distances from the main complex. These objects have $g_{475}-I_{814}$ colors of between 
0.1 and 0.7 magnitudes and $I_{814}$ magnitudes of between 24.3 and 27.7, consistent with star 
formation at rates between 0.5 and 26 $M_\odot$ yr$^{-1}$ \citep{mad1998}.

To determine whether there is a concentration of such objects around the radio galaxy, 
we analyzed the statistics of tadpole and chain galaxies having $23<I_{814}<27$ 
in the whole 3\arcmin$\times$5\arcmin\ ACS field around MRC 1138--262. 
To minimize systematic effects, the morphologies were classified manually by 
somebody not previously associated with the project (E. H.). 
In the 10\arcsec$\times$10\arcsec\ area around the radio galaxy, three such objects were 
found (objects 1, 3, and 4 in Fig. 1). This should be compared with an expected number of 0.22 from 
our analysis of the whole field, a value consistent with the HUDF analysis of \citet{elm2005}. Taking into account the clustering of 
these objects, the probability of finding three such galaxies within a 10\arcsec$\times$10\arcsec\  
region was estimated to be parts in 1000. This implies that the 
chain and tadpole galaxies are concentrated at the position of the radio galaxy and connected with the forming massive galaxy at the 
protocluster centre. 

3. Faint diffuse emission is visible between the obvious clumps. This extended emission is unlikely to be dominated by scattered light of an obscured nuclear quasar, because its morphology is not reminiscent of a scattering cone. Furthermore, its mean color is comparable to that of the star-forming clumps, consistent with the occurrance of ongoing star formation over the whole central 50$\times$40 kpc region. The total extended luminosity (comprising 45\% of the total emission in $g_{475}$) implies a star formation rate of $>$100 $M_\odot$ yr$^{-1}$.

4. Ly$\alpha$ was detected from all of the satellite galaxies in the halo that have 
been studied spectroscopically. The width of the Ly$\alpha$ profile in the halo is consistent with the 
observed velocity dispersion of the associated protocluster, indicating that the galaxies are moving relative to 
each other with radial velocities of up to a few thousand kilometers per second.

5. Although the optical structure is extended approximately along the radio axis, 
most of the galaxy is located outside the narrow region occupied by the radio 
source and is therefore unlikely to be influenced directly by the radio 
source.

\section{Discussion}

\subsection{Evolution of dominant cluster galaxies}

The obvious interpretation of the new {\it HST} image is that it shows 
hierarchical merging processes occurring in a forming massive cD galaxy. 
The morphological complexity and clumpiness observed in the MRC 1138-262 system 
agrees qualitatively with predictions of hierarchical galaxy formation models
\citep[e.g.][]{lar1992,dub1998,gao2004,spr2005a}. 
Because of its striking appearance and its probable nature, we name the host 
galaxy of MRC 1138--262 the ``Spiderweb galaxy'' 
and refer to the small satellite galaxies located within and around it as ``flies.'' 
With relative velocities of several hundred kilometers per second (Fig. 1), these flies 
will traverse the 100 kpc extent of the Spiderweb many times in the 
interval between $z\sim2.2$ and 0, consistent with the 
merger scenario. 
Vigorous merging also provides a plausible mechanism for fueling the supermassive black hole, 
which may subsequently quench any ongoing star formation through radio feedback \citep[e.g.][]{cro2006}. 
Using recent infrared spectroscopic observations of the gas in MRC 1138-262, \citet{nes2006} show that 
the pressure of the radio source is sufficiently large to expel $\sim 50 \%$ of the gas during the 
radio source lifetime. After the gas available for star formation has been expelled, the growth of such 
a massive galaxy will proceed primarily through merging \citep{del2006}.

Assuming that the flies are undergoing single $\sim$1 Gyr starbursts, 
their star formation rates imply a final stellar 
mass for each satellite galaxy of several times $10^{9}$ $M_{\sun}$  and a total mass for all the flies 
of $\sim5\times10^{10}$ $M_{\sun}$. This is less than a tenth of the mass of the whole 
galaxy, derived from its $K$-band luminosity \citep{pen1998}, implying that 
a large fraction of the galaxy mass has already assembled by $z\sim 2.2$. 
Such a relatively early growth disagrees with the predictions of some models for typical evolution of the dominant galaxies in the most massive clusters \citep{del2006}. 

\subsection{Nature of chain and tadpole galaxies}

The most intriguing feature in our {\it HST} image is the association of the linear flies (chains and tadpole morphologies) with the merging massive host galaxy. 
In the HUDF the
frequency of such objects increases dramatically at faint magnitudes \citep[$i_{775}$ $>$ 24;]{elm2005}. 
Because such peculiar galaxies dominate the faint resolved 
galaxy population, they are likely to be an important source of star formation in the early Universe. 
The nature of these objects is unclear. There are several possibilities.
1. They may be spiral galaxies observed edge-on \citep{elm2005}. However, it is difficult to account for the large numbers of faint linear galaxies observed in the HUDF by such selection effects. 

2. Their elongated appearance may be due to star formation associated with radio jets \citep{ree1989} produced by primeval massive black holes \citep{sil1998}. Radio synchrotron jets are known to occur on varying scales, ranging 
from the most luminous galactic nuclei to X-ray binaries and jets are sometimes associated with star formation \citep{vanb1985,bic2000}.

3. They may be formed as the result of merging, either of galaxies that are formed along filamentary gravitational instabilities \citep{tan2001} or in major events $\sim$0.7 Gyr after the merging process has commenced \citep{spr2005b,dim2005,str2006}. 

The fact that the Spiderweb linear flies are located in an environment where vigorous galaxy interactions are taking place 
is consistent with a merger hypothesis for their origin. 
The motions of the flies with velocities of 
several hundred kilometers per second through the dense gaseous halo, 
perturbed by superwinds from the nucleus \citep{zir2005} and the radio jet, 
could result in shocks. The shocks would then lead to Jeans-unstable clouds, 
enhanced star formation along the direction of motion, and chain and tadpole morphologies. 

\section{Conclusions}

The morphology of the Spiderweb galaxy provides a unique new testbed for simulations 
of forming massive galaxies at the centers of galaxy clusters. The occurrence of 
tadpole and chain galaxies in the dense central environment of the protocluster 
places constraints on (1) evolution models for dominant cluster galaxies and 
(2) the nature of the chains and tadpoles, an important constituent of the early Universe.
Our results are consistent with a merger scenario for the formation of these peculiar linear galaxies. 

Deep observations with the {\it HST} of similar objects over a range 
of redshifts are needed to study whether linear flies are generally present 
in the vicinity of cD galaxy progenitors and how their luminosity functions 
in these special regions compare with those of other types of galaxies. The 
positions and morphological parameters of flies in Spiderweb galaxies at $z>2$ 
together with Hubble data on dominant cluster galaxies at $z\sim1$  
will provide new constraints for models of massive galaxy formation. Future 
spectroscopic observations will delineate the velocity field and color 
distributions of MRC 1128--262 in more detail, thereby elucidating further 
how the flies are being captured by the Spiderweb galaxy.

\acknowledgments

G. K. M. acknowledges support from the Royal Netherlands Academy for Arts and 
Sciences and the Netherlands Organization for Scientific Research (NWO). 
J. K. was supported by DFG/ SFB 439. The ACS was developed under NASA contract NAS 5-32864. The research has 
been supported by NASA grant NAG5-7697 and an equipment grant from Sun 
Microsystems, Inc. This Letter is based partially on observations made with 
(1) the ACS on the 
NASA/ESA Hubble Space Telescope, obtained via the STScI, which is operated  
by the AURA Inc., under NASA contract NAS 5-26555, (2) the VLT at 
ESO, Paranal, Chile, program P63.O-0477(A), and (3) the VLA operated by 
the NRAO, which is operated by the AUI. 


\end{document}